\documentstyle[12pt,aps]{revtex}

\def\nc{\newcommand}
\def\nn{\nonumber}
\def\p{\partial}
\nc{\pr}{\frac{\p}{\p r}}
\nc{\pv}{\frac{\p}{\p v}}
\nc{\pta}{\frac{\p}{\p \theta}}
\nc{\pvi}{\frac{\p}{\p \varphi}}

\nc{\pdr}{\frac{\p^2}{\p r^2}}
\nc{\pdv}{\frac{\p^2}{\p v^2}}
\nc{\pdta}{\frac{\p^2}{\p \theta^2}}
\nc{\pdvi}{\frac{\p^2}{\p \varphi^2}}
\nc{\pdva}{\frac{\p^2}{\p v \p \varphi}}
\nc{\pdvr}{\frac{\p^2}{\p v \p r}}
\nc{\pdri}{\frac{\p^2}{\p r \p \varphi}}

\nc{\spr}{\frac{\p}{\p r_*}}
\nc{\spv}{\frac{\p}{\p v_*}}
\nc{\spta}{\frac{\p}{\p \theta_*}}

\nc{\spdr}{\frac{\p^2}{\p r_*^2}}
\nc{\spdvr}{\frac{\p^2}{\p r_* \p v_*}}
\nc{\spdra}{\frac{\p^2}{\p r_* \p \theta_*}}
\nc{\spdri}{\frac{\p^2}{\p r_* \p \varphi}}

\nc{\sta}{\sin\theta}
\nc{\cta}{\cos\theta}
\nc{\sda}{\sin^2\theta}
\nc{\cda}{\cos^2\theta}
\nc{\coa}{\cot\theta}
\nc{\sqd}{\sqrt{2}}
\nc{\cD}{\cal D}
\nc{\cL}{\cal L}
\nc{\cLd}{{\cal L}^{\dagger}}
\nc{\drH}{\dot{r}_H}
\nc{\ddrH}{\ddot{r}_H}
\nc{\prH}{r_H^{\prime}}
\nc{\ka}{\kappa}
\nc{\tkr}{2\kappa(r-r_H)}
\nc{\pprH}{r_H^{\prime\prime}}

\begin{document}
\baselineskip 18pt

\title{Addendum: Hawking Radiation of Photons in a Variable-mass Kerr
Black Hole}
\author{S. Q. Wu\thanks{E-mail: sqwu@iopp.ccnu.edu.cn} and
X. Cai\thanks{E-mail: xcai@ccnu.edu.cn}}
\address{Institute of Particle Physics, Hua-Zhong
Normal University, Wuhan 430079, P.R. China}
\maketitle
\bigskip

\begin{abstract}
Hawking evaporation of photons in a variable-mass Kerr space-time
is investigated by using a method of the generalized tortoise
coordinate transformation. The blackbody radiant spectrum of photons
displays a new spin-rotation coupling effect obviously dependent
on different helicity states of photons.
\end{abstract}

\newpage
Ever since Hawking \cite{Hawk} initiated the discussion on particle
creation by a black hole event horizon, much work has been done on
the Hawking evaporation of black holes in some spherically symmetric
and non-static space-times \cite{Kim,Zhaoel,MY}. In the case of a
non-stationary axisymmetric black hole, though the Hawking radiation
of scalar particles received a fairly extensive investigation \cite{KGNK},
similar work to that of Dirac particles encounters great obstacles.
The main difficulty lies in the non-separability of the radial and
angular variables for Chandrasekhar-Dirac equations \cite{Ch} in the
non-stationary axisymmetry space-time. In a recent paper \cite{WC1}
(here refer to Paper I), this dilemma has been cast off by considering
simultaneously the asymptotic behaviors of the first-order and second-order
forms of Dirac equation near the event horizon. A new interaction due
to the coupling of the spin of Dirac particles with the rotation of the
evaporating Kerr black holes was observed in the thermal radiation spectrum
of Dirac particles. The character of this spin-rotation coupling effect
is its obvious dependence on different helicity states of particles with
spin-$1/2$. This effect vanishes \cite{WC2} when the space-time degenerates
to a spherically symmetric black hole of Vaidya-type.

In this addendum, we apply the method presented in Paper I to deal with
the thermal radiation of photons in a non-stationary Kerr space-time
\cite{GHJW,CKC}, that is, we consider the asymptotic behaviors of the
first-order and second-order forms of Maxwell equations near the event
horizon. By using the relations between the first-order derivatives of
three complex Newman-Penrose \cite{NP} scalars of Maxwell fields, we
eliminate the crossing-terms of the first-order derivatives in the
second-order equation and recast each second-order equation to a standard
wave equation near the event horizon. The blackbody radiation spectrum of
photons displays a spin-rotation coupling effect due to the interaction
between the spin of photons and the angular momentum of radiating Kerr
black holes.

As in Paper I, the line element of a variable-mass Kerr black hole
\cite{GHJW,CKC} can be written in the advanced Eddington-Finkelstein
system as
\begin{eqnarray}
&ds^2& = \frac{\Delta-a^2\sda}{\Sigma}dv^2 +2\frac{r^2+a^2
-\Delta}{\Sigma}a\sda dvd\varphi -2dvdr  \nn\\
&&+2a\sda drd\varphi -\Sigma d\theta^2 -\frac{(r^2+a^2)^2
-\Delta a^2\sda}{\Sigma}\sda d\varphi^2 \, , \label{NKbh}
\end{eqnarray}
where $\Delta=r^2-2M(v)r+a^2$, $\Sigma=r^2+a^2\cda=\rho^*\rho$,
$\rho^*=r+ia\cta$, $\rho=r-ia\cta$, and $v$ is the standard
advanced time. The mass $M$ of the hole depends on the time
$v$, but the specific angular momentum $a$ is a constant.

The metric (\ref{NKbh}) of an evaporating Kerr black hole is a
natural non-stationary generalization of the stationary Kerr
solution. The geometry of this Petrov type-II space-time is
characterized by three kinds of surfaces of particular interest:
the apparent horizons $r_{AH}^{\pm} = M \pm (M^2-a^2)^{1/2}$,
the timelike limit surfaces $r_{TLS}^{\pm} = M \pm (M^2-a^2\cda)^{1/2}$,
and the event horizons $r_{EH}^{\pm} = r_H$. The event horizon is
necessary a null-surface $r = r(v,\theta)$ that satisfies the null
hypersurface conditions $g^{ij}\p_i F\p_j F = 0$ and $F(v,r,\theta) = 0$.
The generalized tortoise coordinate transformation (GTCT) method is
an effective one to determine the location of the event horizon of
a dynamic black hole. To illuminate this method, we use it to derive
the event horizon equation. Because the space-time under consideration
is symmetric about $\varphi$-axis, we can introduce a GTCT \cite{KGNK,WC1}
as follows
\begin{eqnarray}
&&r_*=r +\frac{1}{2\ka(v_0,\theta_0)}\ln[r-r_H(v,\theta)]
\, , \nn \\
&&v_*=v-v_0 \, , ~~~~~~~~~~\theta_*=\theta -\theta_0 \, ,
\label{trans}
\end{eqnarray}
where $r_H = r(v,\theta)$ is the location of event horizon, and $\ka$ is
an adjustable parameter. All parameters $\kappa$, $v_0$ and $\theta_0$
characterize the initial state of the hole and are constant under the
tortoise transformation.

Now applying the GTCT of Eq. (\ref{trans}) to the null surface equation
$$g^{ij}\p_i F\p_j F = 0$$
and then taking the $r \rightarrow r_H(v_0,\theta_0)$, $v \rightarrow v_0$
and $\theta \rightarrow \theta_0$ limits, we arrive at
\begin{equation}
\Big[\Delta_H -2(r_H^2 +a^2)\drH +a^2\sda_0 {\drH}^2
+{\prH}^2\Big]\Big(\spr F\Big)^2 = 0 \, ,
\end{equation}
in which the vanishing of the coefficient in the square bracket can
give the following equation to determine the location of the event
horizon of an evaporating Kerr black hole
\begin{equation}
\Delta_H -2(r_H^2 +a^2)\drH
+a^2\sda_0 {\drH}^2 +{\prH}^2 = 0 \, ,
\label{loeh}
\end{equation}
where we denote $\Delta_H = r_H^2 -2Mr_H +a^2$. The quantities $\drH
=\p r_H/\p v$ is the rate of the event horizon varying in time, $\prH
=\p r_H/\p \theta$ is its rate changing with the angle $\theta$. They
describe the evolution of the black hole event horizon in the time and
the change in the direction, which reflect the presence of quantum
ergosphere near the event horizon. Eq. (\ref{loeh}) is exactly Eq. (11)
derived in Paper I.

Now we consider the sourceless Maxwell equations in the spacetime
(\ref{NKbh}). When the back reaction of the massless spin-$1$ field
on this geometry is neglected, the field equation is given by the
Maxwell equations on the background spacetime (\ref{NKbh}). The sourceless
Maxwell equations in the Newman-Penrose formalism \cite{Ch,NP} read
\begin{eqnarray}
&&(D -2\tilde{\rho})\phi_1 -(\overline{\delta} +\tilde{\pi}
-2\alpha)\phi_0 = -\tilde{\kappa}\phi_2 \, , \nn\\
&&(\delta -2\tau)\phi_1 -(\underline{\Delta} +\mu -2\gamma)\phi_0
 = -\sigma\phi_2 \, , \nn\\
&&(D +2\epsilon -\tilde{\rho})\phi_2 -(\overline{\delta} +2\tilde{\pi})
\phi_1 = -\tilde{\lambda}\phi_0 \, , \nn\\
&&(\delta +2\beta -\tau)\phi_2 -(\underline{\Delta} +2\mu)\phi_1
= -\tilde{\nu}\phi_0 \, . \label{ME}
\end{eqnarray}
To write out their explicit form in the spacetime (\ref{NKbh}), we assume
the complex null tetrad system established in Paper I and insert for the
appropriate spin-coefficients in Eq. (A4) given in Paper I and then after
substituting
$\Phi_0=\frac{\rho^*}{\sqd\rho}\phi_0$,
$\Phi_1=\rho^*\phi_1$, $\Phi_2=\sqd\Sigma\phi_2$,
into Eq. (\ref{ME}), we obtain
\begin{eqnarray}
\Big(\pr +\frac{1}{\rho^*}\Big)\Phi_1
+\Big({\cL}_1 +\frac{ia\sta}{\rho^*}\Big)\Phi_0
&=& 0 \, , \nn \\
\Delta\Big({\cD}_1 -\frac{1}{\rho^*}\Big)\Phi_0
-\Big({\cLd}_0 -\frac{ia\sta}{\rho^*}\Big)\phi_1
&=& 0 \, , \nn \\
\Big(\pr -\frac{1}{\rho^*}\Big)\Phi_2
+\Big({\cL}_0 -\frac{ia\sta}{\rho^*}\Big)\Phi_1
&=& 0 \, ,\nn \\
\Delta\Big({\cD}_0 +\frac{1}{\rho^*}\Big)\Phi_1
-\Big({\cLd}_1 +\frac{ia\sta}{\rho^*}\Big)\Phi_2
&=& 2\dot{M}ria\sta\Phi_0 \, , \label{MP}
\end{eqnarray}
here we have defined operators
\begin{eqnarray*}
&{\cD}_n&=\pr +\frac{2}{\Delta}\Big[n(r-M) +a\pvi +(r^2+a^2)\pv\Big] \, , \\
&{\cL}_n&=\pta +n\coa -\frac{i}{\sta}\pvi -ia\sta\pv \, , \\
&{\cLd}_n&=\pta +n\coa +\frac{i}{\sta}\pvi +ia\sta\pv \, .
\end{eqnarray*}
Eq. (\ref{MP}) can not be decoupled except in the case of
a stationary Kerr black hole \cite{Ch}  ($M= const$). However,
to deal with the problem of Hawking radiation, one should
be concerned about the asymptotic behavior of Eq. (\ref{MP})
near the horizon only. To this end, we consider simultaneously
the asymptotic behaviors of the first-order and second-order
Maxwell equations near the event horizon.

First let us consider the limiting form of Eq. (\ref{MP}) near
the event horizon. Under the transformations (\ref{trans}), Eq.
(\ref{MP}) can be reduced to the following forms
\begin{eqnarray}
&&\Big[\Delta_H -2(r_H^2+a^2) \drH \Big]\spr \Phi_0
+\Big(\prH +ia \sta_0 \drH \Big)\spr \Phi_1  = 0 \, , \nn\\
&&\spr \Phi_1 -\Big(\prH -ia \sta_0 \drH \Big) \spr \Phi_0 = 0 \, ,
\nn\\
&&\Big[\Delta_H -2(r_H^2+a^2) \drH \Big]\spr \Phi_1
+\Big(\prH +ia \sta_0 \drH \Big)\spr \Phi_2  = 0 \, , \nn\\
&&\spr \Phi_2 -\Big(\prH -ia \sta_0 \drH \big) \spr \Phi_1 = 0 \, ,
\label{rela}
\end{eqnarray}
after being taken limits $r \rightarrow r_H(v_0, \theta_0)$,
$v \rightarrow v_0$ and $\theta \rightarrow \theta_0$.

If the derivatives $\spr \Phi_0$, $\spr \Phi_1$ and $\spr \Phi_2$
in Eq. (\ref{rela}) are not equal to zero, the existence condition
of nontrial solutions for $\Phi_0$, $\Phi_1$ and $\Phi_2$ is that
the determinant of two pairs of Eq. (\ref{rela}) vanishes, which
gives exactly the event horizon equation (\ref{loeh}). It will be
seen that the relations (\ref{rela}) play a crucial role to eliminate
the crossing-term of the first-order derivatives in the second-order
equation.

Next we turn to the second-order form of Maxwell equations.
A tedious but straightforward calculation gives
\begin{eqnarray}
&&\Big(\pr\Delta {\cD}_1 +{\cLd}_0{\cL}_1
+2\rho\pv\Big)\Phi_0 = 0 \, , \nn\\
&&\Big(\pr\Delta {\cD}_0 +{\cLd}_1{\cL}_0
-2\rho\pv +\frac{2M\rho}{\rho^{*2}} \Big)\Phi_1
\nn\\
&&~~~~\equiv \Big(\Delta {\cD}_1\pr +{\cL}_1{\cLd}_0 +2\rho\pv
+\frac{2M\rho}{\rho^{*2}}\Big)\Phi_1 \nn\\
&&~~~~~~ = 2ia\sta\Big[\dot{M}r(\pr -\frac{1}{\rho^*})
+\dot{M}\Big]\Phi_0 \, , \nn\\
&&\Big(\Delta {\cD}_0\pr +{\cL}_0{\cLd}_1
-2\rho\pv\Big)\Phi_2 \nn\\
&&~~~~~~ = -2\ddot{M}ra^2\sda\Phi_0
-4\dot{M}ria\sta{\cL}_1\Phi_0 \, ,
\label{foe}
\end{eqnarray}

Given the GTCT in Eq. (\ref{trans}), the limiting form of Eq.
(\ref{foe}), when $r$ approaches $r_H(v_0, \theta_0)$, $v$ goes
to $v_0$ and $\theta$ goes to $\theta_0$, reads
\begin{eqnarray}
&&\Bigg\{\Big[\frac{r_H(1 -2\drH) -M}{\ka} +2\Delta_H
-2\drH (r_H^2+a^2)\Big]\spdr -2\prH \spdra  \nn \\
&&+2a(1 -\drH) \spdri +2(r_H^2 +a^2 -\drH a^2\sda_0)
\spdvr -\Big[2(M-r_H) \nn \\
&& +2(r_H-ia\cta_0)\drH +\prH \coa_0 +\pprH
+\ddrH a^2\sda_0\Big]\spr\Bigg\} \Phi_0 = 0 \, ,  ~~~~
\label{wzero}
\end{eqnarray}
and
\begin{eqnarray}
&&\Bigg\{\Big[\frac{r_H(1 -2\drH) -M}{\ka}+ 2\Delta_H
-2\drH (r_H^2+a^2)\Big]\spdr -2\prH \spdra  \nn \\
&&+2a(1 -\drH) \spdri +2(r_H^2 +a^2 -\drH a^2\sda_0)
\spdvr   \nn \\
&&-\Big(-2r_H\drH +\prH \coa_0 +\pprH
+\ddrH a^2\sda_0\Big)\spr\Bigg\} \Phi_1 \nn \\
&& = -2\dot{M}r_H ia\sta_0 \frac{\prH
+ia \sta_0 \drH}{\Delta_H -2(r_H^2+a^2)\drH} \spr\Phi_1 \, ,
\label{wone}
\end{eqnarray}
and
\begin{eqnarray}
&&\Bigg\{\Big[\frac{r_H(1 -2\drH) -M}{\ka}+ 2\Delta_H
-2\drH (r_H^2+a^2)\Big]\spdr -2\prH \spdra  \nn \\
&&+2a(1 -\drH) \spdri +2(r_H^2 +a^2 -\drH a^2\sda_0)
\spdvr -\Big[2(r_H -M)  \nn \\
&&+2(-3r_H+i a\cta_0)\drH +\prH \coa_0
+\pprH +\ddrH a^2\sda_0\Big]\spr\Bigg\} \Phi_2 \nn \\
&& = -4\dot{M}r_H ia\sta_0 \frac{\prH
+ia \sta_0 \drH}{\Delta_H -2(r_H^2+a^2)\drH} \spr\Phi_2 \, , ~~~~
\label{wtwo}
\end{eqnarray}
where we have replaced the first-order derivative term
$\spr \Phi_0$ in Eqs. (\ref{wone}, \ref{wtwo}) by using
the relations (\ref{rela}). In the calculations, the
L'H\^osptial's rule has been used to treat an infinite
form of $0/0$-type.

In order to reduce Eqs. (\ref{wzero}), (\ref{wone}) and
(\ref{wtwo}) to a standard form of wave equation near the
event horizon, we adjust the parameter $\ka$ as did in Paper
I, then these wave equations can be recast into an united form
as follows
\begin{equation}
\Big[\spdr +2\spdvr +2\Omega \spdri
+2C_3 \spdra +2(C_2 +iC_1) \spr\Big] \Psi_p = 0 \, ,
\label{wave}
\end{equation}
where the angular velocity of the event horizon of the evaporating
Kerr black hole, $\Omega$, and the coefficient $C_3$ are presented in
Paper I, while both $C_1$ and $C_2$ are real,
\begin{eqnarray*}
C_2+iC_1&=&\frac{-1}{2(r_H^2 +a^2 -\drH a^2\sda_0)}
\Big[2p(r_H-M+ia\cta_0\drH) \\
&-&2(2p+1)r_H\drH +\prH \coa_0 +\pprH +\ddrH a^2\sda_0  \\
&+&2(s+p)\dot{M}r_Hia\sta_0\frac{\prH +ia\sta_0\drH}{\Delta_H
-2\drH (r_H^2+a^2)} \Big] \, ,  \\
\end{eqnarray*}
in which the correspondence should be $\Psi_p=\Phi_0$, $\Phi_1$,
$\Phi_2$ when $p=-1, 0, 1$ for ($s=1$), respectively. It should
be pointed out that Eq. (\ref{wave}) includes, as a special case,
the wave equation (19) of Dirac particles discussed in Paper I,
that is, $\Psi_p=P_2$, $P_1$, when $p=-1/2, 1/2$ for ($s=1/2$).

The subsequent analysis parallels the treatment of Paper I, and
one can obtain the Hawking radiation spectrum of photons from the
black hole by using the method of Damour-Ruffini-Sannan's \cite{DRS},
\begin{equation}
\langle {\cal N}_{\omega} \rangle \sim
\frac{1}{e^{(\omega -m\Omega - C_1)/T} -1} \, ,
~~~~ T=\frac{\ka}{2\pi} \, . \label{sptr}
\end{equation}
where $\omega$ is the energy of photons, $m$ is the azimuthal quantum
number. The explicit expression of the surface gravity $\ka$ is referred
to Paper I.

The thermal radiation spectrum (\ref{sptr}) due to the Bose-Einstein
statistics of photons shows that the black hole emits radiation just
like a black body. It demonstrate that the energy spectrum of photons
in an evaporating Kerr space-time is composed of two parts:
\begin{eqnarray}
\omega_p &=& \frac{a}{r_H^2 +a^2 -\drH a^2\sda_0}
\Big[m(1-\drH) -p\cta_0\drH  \nn\\
&& +(s+p)\dot{M}r_H\frac{\sta_0\prH}{\Delta_H
-2\drH(r_H^2+a^2)}\Big] \, ,
\end{eqnarray}
the rotational energy $m\Omega$ arising from the coupling of the orbital
angular momentum of photons with the rotation of the black hole and $C_1$
due to the coupling of the intrinsic spin of photons and the angular
momentum of the black hole. From the explicit expression of the \lq\lq
spin-dependent" term $C_1$
\begin{equation}
C_1 = \frac{\Omega}{1-\drH}\Big[-p\cta_0\drH
+(s +p)\dot{M}r_H\frac{\sta_0\prH}{\Delta_H
-2\drH \big(r_H^2+a^2\big)}\Big] \, ,
\end{equation}
one can easily find that it vanishes in the case of a stationary
Kerr black hole ($M = const$, $\drH = \prH = 0$) or a Vaidya-type
black hole ($a = 0$, $\prH = 0 $, $\drH \not= 0$). The term $C_1$
is obviously related to the helicity of photons in different spin
states, it characterizes a new effect arising from the interaction
between the spin of photons and the rotation of an evaporating black
hole.

In this addendum, we have dealt with Hawking radiation of photons in
a variable-mass Kerr black hole. Eq. (\ref{sptr}) shows the thermal
radiation spectrum of photons in the non-stationary Kerr space-time,
in which an extra term $C_1$ represents a new spin-rotation coupling
effect probably arising from the interaction between the spin of photons
and the angular momentum of the evaporating Kerr black hole. The
feature of this spin-rotation coupling effect is its dependence on
different helicity states of photons. As is pointed out in Paper I,
this effect vanishes when a black hole is stationary ($\drH = \prH = 0$)
or it has a zero angular momentum ($a = 0$).

In summary, this study confirms that the thermal radiation spectra of
particles with higher spins in the non-stationary Kerr black hole displays
a new effect due to the coupling of the intrinsic spin of particles with
the rotation of the black holes. The spin-rotation coupling effect shows
that a non-stationary Kerr space-time has some distinct effects
different from that of a stationary Kerr black hole. This effect
is not shared by a stationary Kerr black hole or a Vaidya-type
spherically symmetric black hole.

This work is supported in part by the NSFC in China.

\end{document}